\documentclass[
    twocolumn,
	prd,
	amssymb,
	preprintnumbers,
	secnumarabic,
	nofootinbib,
	superscriptaddress]{revtex4-1}

\pdfoutput=1

\usepackage{graphicx}
\usepackage{enumitem}
\usepackage{latexsym}
\usepackage{amsfonts}
\usepackage{amssymb}
\usepackage{color}
\usepackage{amsmath}
\usepackage{slashed}
\usepackage{dcolumn}
\usepackage{verbatim}
\usepackage{float}
\usepackage{multirow}
\usepackage{xspace}
\usepackage[normalem]{ulem}
\usepackage{hyperref}

\definecolor{greeny}{rgb}{0.3,0.7,0.3}

\definecolor{hypershade}{rgb}{0.8,0.3,0.3}
\hypersetup{
  pdfauthor={Nirmal Raj},
  pdftitle={Pulsating pulsars},
  pdfsubject={Pulsating pulsars},
  colorlinks=true,
  citecolor=magenta,
  urlcolor=blue,
  linkcolor=hypershade
}

\newcommand{\lsim}{\lesssim}

\def\Oc{\mathcal{O}}

\newcommand{\acro}[1]{\textsc{\MakeLowercase{#1}}} 

\renewcommand{\tilde}{\widetilde} 

\newcommand{\beq}{\begin{equation}}
\newcommand{\eeq}{\end{equation}}
\newcommand{\bea}{\begin{eqnarray}}
\newcommand{\eea}{\end{eqnarray}}
\newcommand{\nn}{\nonumber}

\def \Rsh {R_{\rm sh}}
\def \Msh {M}
\def \mdm {m_\chi}

\def \vdm {\sigma_v}
\def \vesc {v_{\rm esc}}
\def \rhodm {\rho_{\chi}}
\def \fdm {f_\chi}

\def \RNS {R_{\rm NS}}
\def \rco {R_{\rm co}}
\def \MNS {M_{\rm NS}}

\def \TNSz {\tilde T_{\rm NS}}

\def \tmeet {t_{\rm meet}}
\def \tcool {t_{\rm cool}}

\def \Thotz {\tilde{T}_{\rm hot}}
\def \Tcoldz {\tilde{T}_{\rm cold}}
\def \Tcoolz {\tilde{T}_{\rm cool}}

\allowdisplaybreaks

\begin{document}

\title{Scattering searches for dark matter in subhalos:\\ neutron stars, cosmic rays, and old rocks}

\author{Joseph Bramante}
\email{joseph.bramante@queensu.ca}
\affiliation{The Arthur B. McDonald Canadian Astroparticle Physics Research Institute,
Department of Physics, Engineering Physics, and Astronomy,
Queen’s University, Kingston, Ontario, K7L 2S8, Canada}
\affiliation{Perimeter Institute for Theoretical Physics, Waterloo, Ontario, N2L 2Y5, Canada}

\author{Bradley J. Kavanagh}
\email{kavanagh@ifca.unican.es}
\affiliation{Instituto de F{\'i}sica de Cantabria (IFCA, UC-CSIC), Avenida de Los Castros s/n, 39005 Santander, Spain}

\author{Nirmal Raj}
\email{nraj@triumf.ca}
\affiliation{TRIUMF, 4004 Wesbrook Mall, Vancouver, BC V6T 2A3, Canada}

\date{\today}

\begin{abstract}
In many cosmologies dark matter clusters on sub-kiloparsec scales and forms compact subhalos, in which the majority of Galactic dark matter could reside.
Null results in direct detection experiments since their advent four decades ago could then be the result of extremely rare encounters between the Earth and these subhalos.
We investigate alternative and promising means to identify subhalo dark matter interacting with Standard Model particles:
(1) subhalo collisions with old neutron stars can transfer kinetic energy and brighten the latter to luminosities within the reach of imminent infrared, optical, and ultraviolet telescopes; 
we identify new detection strategies involving single-star measurements and Galactic disk surveys, and obtain the first bounds on self-interacting dark matter in subhalos from the coldest known pulsar, PSR J2144\textendash{}3933,
(2) subhalo dark matter scattering with cosmic rays results in detectable effects,
(3) historic Earth-subhalo encounters can leave dark matter tracks in paleolithic minerals deep underground.
These searches could discover dark matter subhalos weighing between gigaton and solar masses, with corresponding dark matter cross sections and masses spanning tens of orders of magnitude.
\end{abstract}

\maketitle

{\bf \em Introduction.}
The enigma of dark matter persists~\cite{Bertone:2004pz}.
In the hope that its microscopic properties will be revealed through its scattering and annihilation into Standard Model (SM) states, extensive experimental efforts are underway to detect it~\cite{Baudis:2014naa,
Gaskins:2016cha,
Schumann:2019eaa}.
Both direct and indirect dark matter (DM) searches often assume that DM is smoothly distributed over the Galactic halo. 
However, cold DM is expected to form subhalo structures down to Earth-mass scales ($\sim 10^{-6}\,M_\odot$)~\cite{Zybin:1999ic,Hofmann:2001bi,Berezinsky:2007qu}. 
Such substructure should persist on length scales below a kiloparsec if small DM halos remain un-disrupted during hierarchical clustering, as suggested by $N$-body simulations~\cite{BergstromGondolo:1998jj,vandenBosch:2017ynq,vandenBosch:2018tyt}.
Furthermore, in many cosmological scenarios the small-scale power is enhanced -- via, e.g., an early matter-dominated era or DM self-interactions augmenting primordial density perturbations  -- resulting in large fractions of DM surviving in subhalos~\cite{ErickcekSigurdson,Barenboim:2013gya,FanWatson,drorcodecay,inflatflucs,Buckley:2017ttd,nussinovcluster,Barenboim:2021swl}, where certain collisionless DM cosmologies predict subhalos masses ranging from $10^{-28}-10^{7} ~M_{\odot}$ \cite{ErickcekSigurdson,Erickcek:2020wzd,Erickcek:2021fsu}.

It is known that subhalos could boost Galactic indirect detection fluxes, since the DM annihilation rate scales as the density squared~\cite{Gondolo:1990dk,Hiroshima:2018kfv,Blanco:2019eij,StenDelos:2019xdk,Delos:2019lik}.
Due attention has thereby been paid to annihilation signals, but it may be that DM only reveals its SM interactions  via scattering processes, such as when it is asymmetric~\cite{Petraki:2013wwa} or its annihilation is $p$-wave-dominated~\cite{Zhao:2016xie,Zhao:2017dln,Boddy:2019qak}.
Should the interacting component of DM reside in subhalos, direct detection (DD) experiments~\cite{Drukier:1984vhf,GoodmanWitten:1984dc,Drukier:1986tm} may have observed no conclusive signal simply because the Earth has yet to encounter a subhalo since their inception.
Thus other, novel probes of DM scattering are needed; in this {\em Letter} we outline three strategies.

The first is to observe heating of old, nearby neutron stars (NSs) by subhalo DM. These NSs will travel through subhalos and capture constituent DM particles, in a process that transfers DM kinetic energy of order the DM mass, because of the steep gravitational potential around NSs.
This method was first proposed to detect smooth halo DM~\cite{NSvIR:Baryakhtar:DKHNS}, which would result in infrared emission from NSs observable by upcoming telescopes, and is currently an active area of research~\cite{snowmass:Berti:2022rwn,NSvIR:Raj:DKHNSOps,NSvIR:Pasta,NSvIR:SelfIntDM,NSvIR:Bell2018:Inelastic,NSvIR:GaraniGenoliniHambye,NSvIR:Queiroz:Spectroscopy,NSvIR:Hamaguchi:Rotochemical,NSvIR:Marfatia:DarkBaryon,NSvIR:Bell:Improved,NSvIR:DasguptaGuptaRay:LightMed,NSvIR:GaraniGuptaRaj:Thermalizn,NSvIR:Queiroz:BosonDM,NSvIR:Bell2020improved,NSvIR:zeng2021PNGBDM,NSvIR:anzuiniBell2021improved,NSvIR:Bell2019:Leptophilic,NSvIR:GaraniHeeck:Muophilic,NSvIR:Riverside:LeptophilicShort,NSvIR:Riverside:LeptophilicLong,NSvIR:Bell:ImprovedLepton,NSvIR:Bramante:2021dyx,NSvIR:Fujiwara:2022uiq,NsvIR:Hamaguchi:2022wpz}.
Here we show that subhalo DM can impart larger NS luminosities than diffuse DM, bringing imminent optical and ultraviolet observations into play.
Moreover, for DM with strong self-interactions, we show that due to enhanced accretion there is parameter space {\em already} constrained by observations of the coldest known NS, PSR J2144\textendash{}3933~\cite{coldestNSHST}.
All in all, NS luminosity searches can probe gigaton to solar mass subhalos, for DM-nucleon cross sections as low as $10^{-45}~{\rm cm}^2$. 

Two other strategies to detect distant DM subhalos involve terrestrial probes that do not rely on Earth's spatial position or the usual years-long runtime of terrestrial experiments.
Specifically, encounters of subhalo DM with Galactic cosmic rays and with geological minerals deep underground can be discerned in several ways discussed below. 
Under the assumption of a smooth halo, cosmic rays and minerals have already been utilized to constrain strongly-interacting DM~\cite{CR:Cyburt:2002uw,CR:BringmannPospelov,CR:Ema:2018bih,CR:NuExpCappiello:2019qsw,CR:PROSPECT:2021awi,Price:1986ky,SnowdenIfft:1995ke};
here we will use these to set limits on DM subhalo masses and radii.

Current constraints on subhalos come from gravitational microlensing surveys, applicable for asteroid to solar mass subhalos with sizes ranging up to stellar radii~\cite{microlens:erosogle,microlens:subaru}, and
additional gravitational probes include pulsar timing arrays~\cite{PTA:Dror:2019twh} and accretion glows in molecular clouds~\cite{dMACHOS:Bai:2020jfm}.
Direct detection of clustered DM up to the Planck mass was studied in Ref.~\cite{nussinovcluster}, of higher mass composite DM in Refs.~\cite{xmaslights:Bramante:2018qbc,xmaslights:Bramante:2018tos,xmaslights:Bramante:2019yss,Acevedo:2020gro,Acevedo:2020avd,xmaslights:DEAP:2021fum,Acevedo:2021kly,snowmass:Carney:2022gse}, and
similar sensitivities with ancient minerals in  Refs.~\cite{Jacobs:2014yca,Ebadi:2021cte,Acevedo:2021tbl}.
Asteroid-mass subhalos colliding with stars were studied in Ref.~\cite{DasEllisSchusterZhou:2021drz}, assuming geometric cross sections. 
We do not make this assumption and treat more general microscopic DM cross sections. 
Finally, DM substructure could impact DD experiments~\cite{DDclumps:WidrowStiff:2001dq,DDclumps:Kamionkowski:2008vw,bradleyclumpSun} and the biosphere~\cite{clumpyXtinxion:Collar:1995as}.

{\bf \em Neutron star heating and cooling.}
In this study we consider subhalos of uniform density of mass $\Msh$ and maximum radius $\Rsh$; other density profiles are expected to impact these results at the $\Oc(1)$ level.
For simplicity we assume all NSs are of the same mass $\MNS$ and radius $\RNS$, with benchmark parameters
\beq
\MNS = 1.5~M_\odot~, \ \ \RNS = 10~{\rm km}~.
\label{eq:NSbenchmark}
\eeq

The rate of NS-subhalo encounters for a single NS with  $\MNS \gg \Msh$ is given by (see Supplementary Material):
\bea
\label{eq:Gammameet}
\nn \Gamma_{\rm meet} (r) =&& 4\sqrt{\pi}\fdm  \frac{\rhodm(r)}{\Msh} \vdm(r)  \times  \\
&& \bigg[\Rsh+\RNS \sqrt{1+\bigg(\frac{\vesc}{\langle v_{\rm rel}\rangle}\bigg)^2} \bigg]^2
\eea
where $\rhodm (r)$ and $\vdm(r)$ are the DM density and speed dispersion at Galactic position $r$, and the escape velocity at the NS surface is $\vesc = \sqrt{2 G \MNS/\RNS}$.  
Here we have averaged the rate over the relative velocity $v_\mathrm{rel}$ of the encounters, which we assume to follow a Maxwell-Boltzmann distribution with dispersion $\sigma_{v_\mathrm{rel}}(r) = \sqrt{2} \vdm$~\cite[Problem 8.8]{BinneyTremaine:2008}.
For the solar position $r_\odot = 8.3$~kpc, we have $\rho_\odot = 0.4$~GeV/cm$^3$,  
$\vdm = 156$~km/s~\cite{vcircMW} and the average subhalo-NS relative speed $\langle v_{\rm rel}\rangle \simeq 350$ km/s,
and for $\Rsh \ll \RNS$: 
\beq
\Gamma_{\rm meet}(r_\odot) = \frac{f_\chi}{1.9~{\rm Gyr}} \bigg( \frac{10^{-15} M_\odot}{\Msh} \bigg)~.
\eeq
For concreteness, in our numerical results we fix the fraction of DM in subhalos as $f_\chi = 1$.

Assuming that all of the incident DM flux in the subhalo is captured by the NS, the DM kinetic energy deposited in the NS after subhalo passage is
\beq
E_{\rm meet} = z \Msh \min \bigg[1, \bigg(\frac{\rco}{\Rsh}\bigg)^2  \bigg]~,
\label{eq:EpassMsh}
\eeq
where the blueshift of DM falling into the neutron star increases the DM kinetic energy according to
\bea
1 + z &=& \frac{1}{\sqrt{1-2G \MNS/\RNS}} = 1.34~, 
\eea
and the effective gravitational collection radius $\rco$ of the neutron star is:
\bea
\rco = \bigg( \frac{\vesc}{\langle v_{\rm rel}\rangle} \bigg)^\ell \RNS (1 + z)~,
\label{eq:rco}
\eea
where
$\ell=1$ if the DM in the subhalo is effectively collisionless~\cite{Goldman:1989nd}, and
$\ell=2$ for DM that behaves like a fluid, in which case $\rco$ will be the Bondi radius \cite{BondiHoyle1944,Bondi1952,Begelman1977,Shima1985}.
The latter expression assumes that the sound speed in the subhalo 
does not exceed the NS-subhalo relative velocity.
In Eq.~\eqref{eq:rco}, 
the $1+z$  factor accounts for the apparent NS radius to distant observers. 
DM in subhalos will heat an NS from internal temperature $\Tcoldz$ to $\Thotz$ given by 
\beq
\int^{\Thotz}_{\Tcoldz} d\tilde T c_V (\tilde T) = E_{\rm meet}~. 
\label{eq:TcVinteg}
\eeq
Here and throughout this paper, the tilde denotes temperatures measured by a distant observer: $\tilde T \equiv T/(1+z)$, where $T$ is the NS local core temperature.

The NS heat capacity $c_V$ is given by~\cite{cvanalytic:Ofengeim:2017xxr} (see Appendix for more on NS cooling)

\bea
\nn c_V (\tilde T) &=& 4.8 \times 10^{26}~{\rm J/K}~\bigg(\frac{\tilde{T}}{10^4~{\rm K}}\bigg) \\
               &=& 2.7 \times 10^{-21}~M_\odot/{\rm K}~\bigg(\frac{\tilde{T}}{10^4~{\rm K}}\bigg)~.
\label{eq:heatcapelec}
\eea

Putting Eqs.~\eqref{eq:TcVinteg} and \eqref{eq:heatcapelec} together, we obtain
\beq
\bigg(\frac{\Thotz}{10^4~{\rm K}}\bigg)^2 = \bigg(\frac{\Tcoldz}{10^4~{\rm K}}\bigg)^2 + \frac{E_{\rm meet}}{6.2 \times 10^{-18}~M_\odot}~.
\label{eq:ThotTcoldEpass}
\eeq

Between subhalo encounters the NS cools by emitting photons from its surface and neutrinos from its interior; the latter mode is sub-dominant for $\tilde T \lsim 10^8$~K, which corresponds to a surface temperature (generally different from $\tilde T$ due to an insulating ``envelope" on the NS) of $\tilde T_s \lsim 10^6$~K.
We take the NS cooling time $\tcool$ from Ref.~\cite{coolinganalytic:Ofengeim:2017cum}; see Appendix for further notes.

{\bf \em Frequent NS encounters.}
If subhalos heat NSs often enough, the NSs will maintain a steady-state temperature. Specifically, if the NS cooling time $\tcool (\tilde T_s)$ exceeds the timescale between subhalo encounters $t_{\rm meet} = \Gamma_{\rm meet}^{-1}$, and if the encounters are energetic enough to impart temperatures $> \tilde T_s$, the NS will maintain a temperature of at least $\tilde T_s$.
To find this temperature in terms of subhalo parameters, we set $\Thotz^2 = 2 \Tcoldz^2$ in Eq.~\eqref{eq:ThotTcoldEpass} to estimate the $E_{\rm meet}$ required to attain a steady-state temperature in the vicinity of $\Tcoldz$, which we call $E^{\rm T}_{\rm meet}$.
Then for a given subhalo radius and NS accretion rate, $E^{\rm T}_{\rm meet}$ corresponds to some $\Msh$ in Eq.~\eqref{eq:EpassMsh}, which also implies a unique $t_{\rm meet}$ in Eq.~\eqref{eq:Gammameet}.

\begin{figure}[t]
    \centering
    \includegraphics[width=0.45\textwidth]{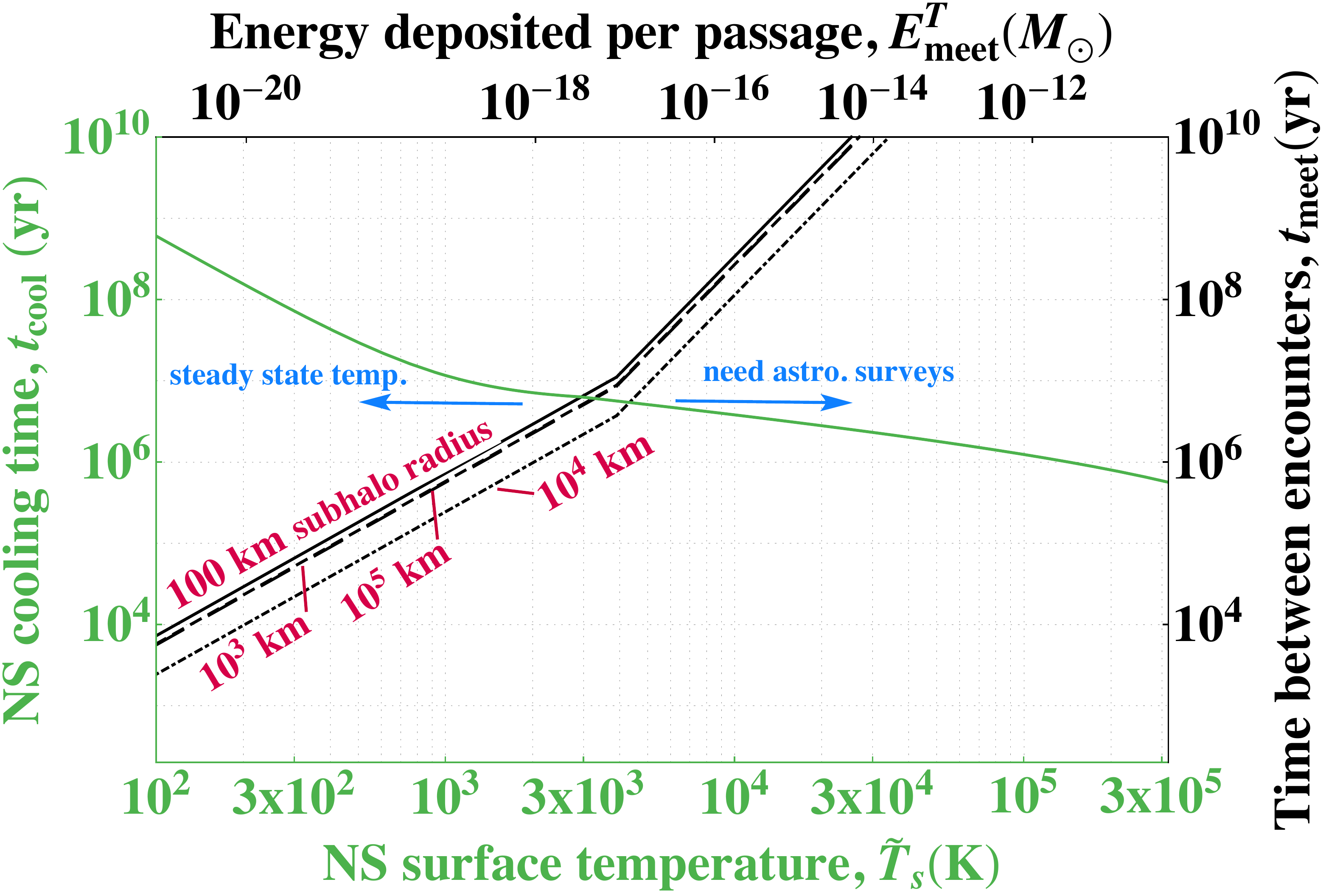}
    \caption{Regions showing ``frequent" vs ``infrequent" NS-subhalo encounters, requiring different observational strategies.
    \textcolor{greeny}{{\bf Green}}: Neutron star cooling time as a function of surface temperature $\tilde T_s$, assuming superfluid nucleons.
    {\bf Black}: The time between encounters at the Solar position for various subhalo radii, as a function of the DM kinetic energy deposited in NSs per encounter,
    obtained from Eqs.~\eqref{eq:Gammameet} and \eqref{eq:EpassMsh}.
    The top x-axis shows energies required to maintain corresponding steady-state $\tilde T_s$ values in the bottom x-axis.
       To the left of black-green intersections, observations of a single NS with indicated temperatures would probe corresponding subhalo masses and radii.
    To the right, astronomical surveys are required to find NS ensembles exceeding indicated temperatures.
    See text for further details.
          }
    \label{fig:tvsTE}
\end{figure}

In Fig.~\ref{fig:tvsTE} we plot as a function of $\tilde T_s$ (corresponding to the $E^{\rm T}_{\rm meet}$ in the top x-axis) both $\tcool$ and $t_{\rm meet} (r_\odot)$ for various $\Rsh$. The condition $\tmeet < \tcool$ then defines the NS temperatures for which a steady steady state scenario is possible.
It can be seen that the subhalo encounter timescale does not exceed the cooling time for $\tilde T_s \lsim 3000$~K in the limits where the subhalo is much smaller or much larger than the NS, 
whereas for $\Rsh$ comparable to $\RNS$ the condition could produce hotter NSs.
This follows from Eqs.~\eqref{eq:Gammameet} and \eqref{eq:EpassMsh}. 
For  $\Rsh \ll \RNS$ both $t_{\rm meet}$ and $E_{\rm meet}$ $\propto \Msh$ , and for $\Rsh \gg \RNS$ both $t_{\rm meet}$ and $E_{\rm meet}$ $\propto \Msh \Rsh^{-2}$, thus the $t_{\rm meet}$ vs $E^{\rm T}_{\rm meet}$ curves are co-incident in these limits.
For intermediate $\Rsh$ the Sommerfeld enhancement factors for $\Gamma_{\rm meet}$ and $E_{\rm meet}$ differ in such a way that greater energies can be deposited in less frequent encounters.

\begin{figure*}
    \centering
    \includegraphics[width=\textwidth]{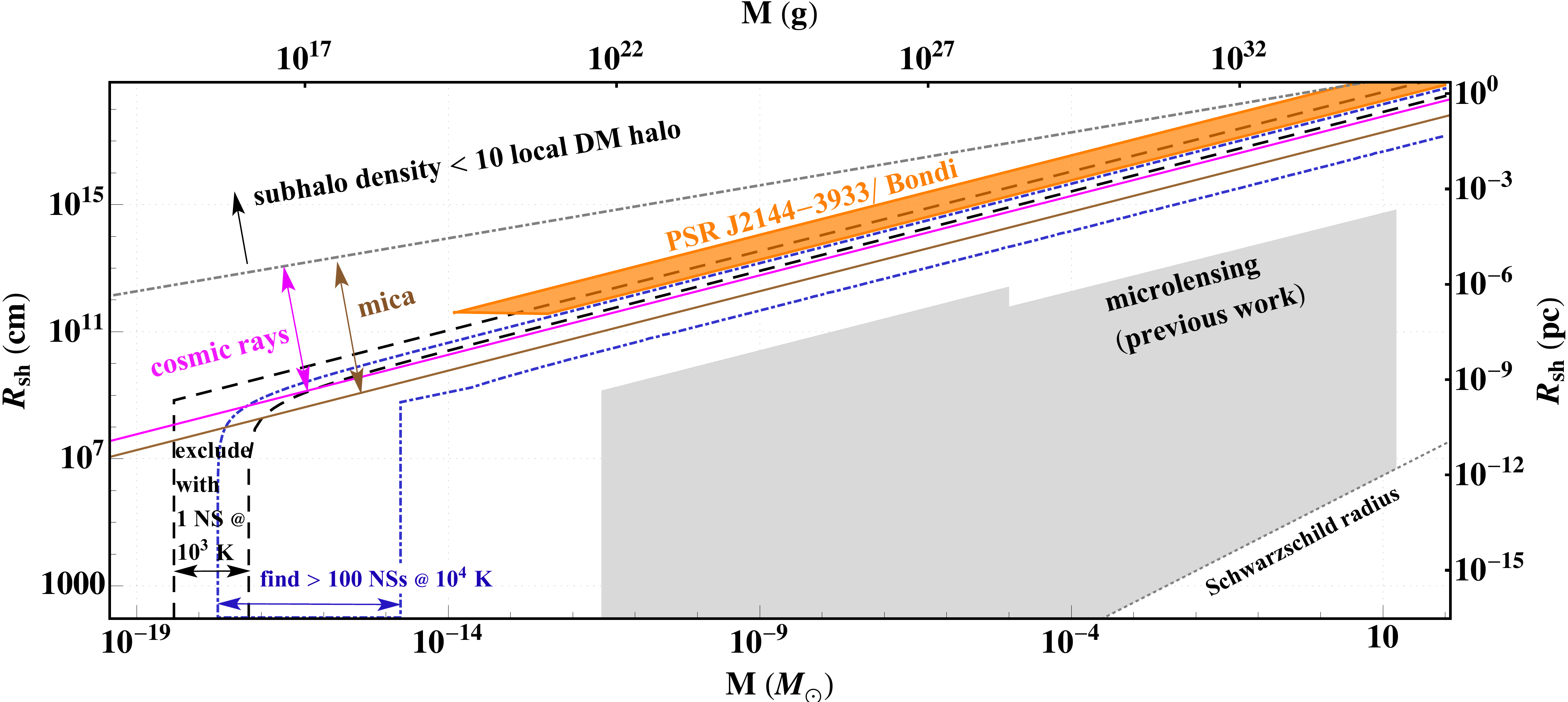} \\
    \includegraphics[width=\textwidth]{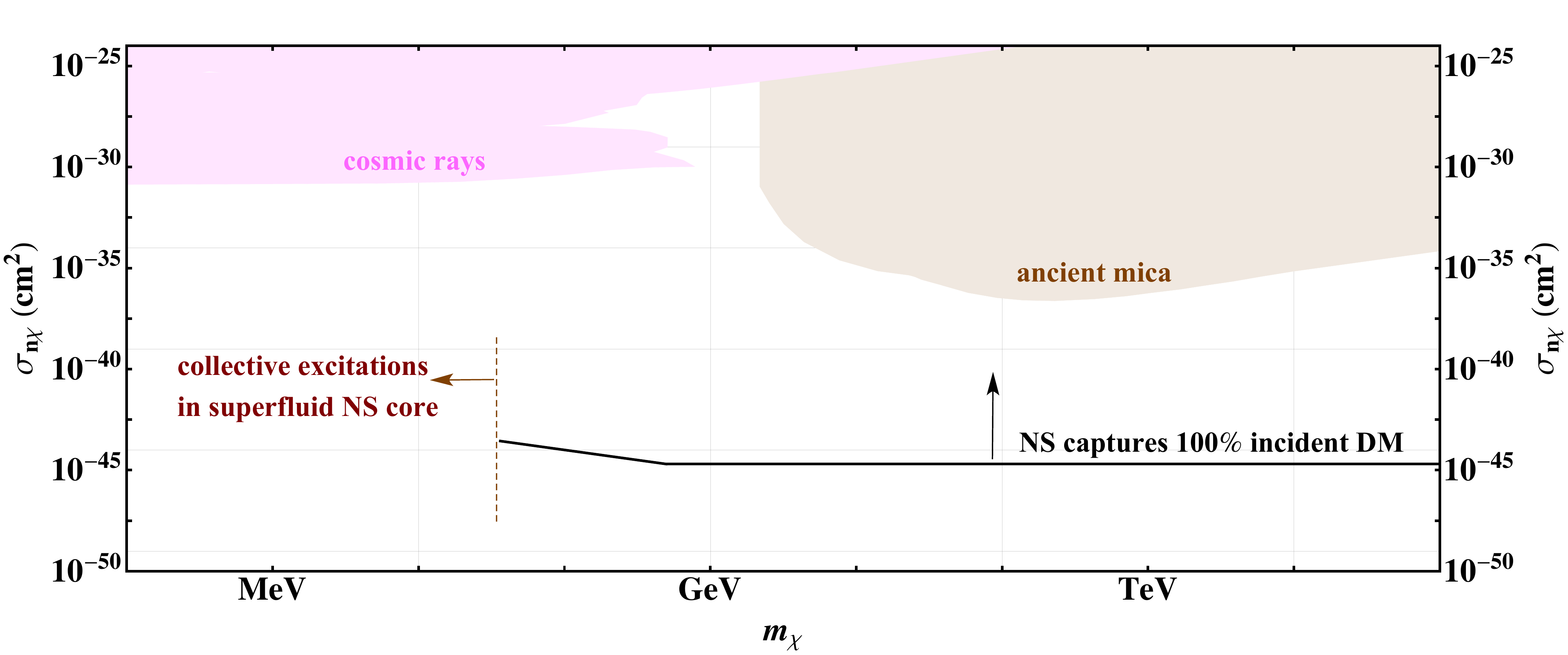}
    \caption{Subhalo masses and radii probed by our detection strategies in the {\bf top} panel, corresponding to limits on DM mass and nucleon scattering cross sections shown in the {\bf bottom} panel.
     For collisionless DM, we show regions that can be probed by the observation of a single NS with 1000 K surface temperature (black dashed lines), as well as by future telescope surveys of NSs within a kpc of Earth at the temperatures indicated (blue dot dashed lines).
    For dissipative/fluid DM, we exclude the orange region using observations of the coldest NS, for cross sections above the ``100\% capture'' line in the bottom panel.
       Also shown are regions corresponding to existing constraints from DM-cosmic ray scattering (pink) and track searches in ancient mica (brown).
    Above the grey dot-dashed line, the galactic DM density exceeds the density in subhalos so DM does not significantly cluster.
    Limits from microlensing surveys are also shown.
        See text for more details.}
     \label{fig:$}
\end{figure*}

The coldest NS observed, PSR J2144\textendash{}3933, has a temperature upper limit of 2.9$\times 10^4$~K for our benchmark NS~\cite{coldestNSHST}, whereas our largest achievable steady-state temperature is $\tilde T_s < 6\times 10^3$~K (for $\Rsh = 10^4$~km), for DM that accretes onto NSs in the collisionless regime, $i.e.$ $\ell=1$ in Eq.~\eqref{eq:rco}. In the top panel of Fig.~\ref{fig:$} we show, in the space of $\Msh$ vs $\Rsh$, the subhalo parameters that can be explored in the future by measuring $\tilde T_s$ down to 1000 K using upcoming infrared telescopes like JWST, ELT and TMT, within achievable times~\cite{NSvIR:Baryakhtar:DKHNS,NSvIR:Raj:DKHNSOps}.
This region is bounded on the left by a black-dashed contour of $\Thotz = 1000$~K (obtained from Eq.~\eqref{eq:ThotTcoldEpass} for $\Tcoldz \ll \Thotz$), and on the right by a black-dashed contour of $t_{\rm meet} = \tcool (1000~{\rm K}) = 1.2 \times 10^7$~yr, 
and would be excluded if an NS at $r = r_\odot$ is measured to be below 1000 K. (For a similar region corresponding to $10^4$~K, see Supplementary Material.) 

On the other hand, if DM has substantial self-interactions then PSR J2144\textendash{}3933 {\em already} places a strong bound on subhalo DM.
In this case we consider fluid accretion of the DM onto NSs, especially relevant for theories of dissipative and strongly self-interacting DM~\cite{Chang:2018bgx,Huo:2019yhk,Liu:2019bqw,Shen:2021frv,Ryan:2021dis}, for which DM can have a sizable sound speed (much like the interstellar medium), if collected into dense subhalos. 
The effective DM collection radius of the NS will now be enhanced by another factor of $\vesc/\left\langle v_\mathrm{rel}\right\rangle$ ($i.e.$ $\ell=2$ in Eq.~\eqref{eq:rco}) \cite{BondiHoyle1944,Bondi1952,Begelman1977,Shima1985}, and in a large parametric region Bondi accretion imparts enough energy for all NSs near Earth to have temperatures in excess of 2.9$\times 10^4$~K.
In Fig.~\ref{fig:$}, the orange region is constrained by the $\tilde T_s \lesssim 3 \times 10^4~{\rm K}$ bound on PSR J2144\textendash{}3933's temperature \cite{coldestNSHST}. As discussed in the Supplemental Material, NS accretion of DM may be enhanced beyond the estimate given here via disk formation if these self-interacting DM subhalos are tidally disrupted, although disruption itself will depend on the binding potential of the subhalo.
In addition, subhalo DM with strong self-interactions could be constrained using $e.g.$ Bullet Cluster observations -- this would require a detailed simulation of subhalo collisions in the Bullet Cluister, and the specific self-interacting DM model~\cite{Tulin:2017ara}.

The regions just discussed may be probed so long as the dynamics of the DM-NS scattering process lead to capture. This condition is depicted in black in the bottom panel of Fig.~\ref{fig:$}, for velocity-independent elastic scattering, modeling the NS as a free fermion gas; in the region above the black curve all the DM flux incident on the NS is captured.
As explained in Refs.~\cite{Goldman:1989nd,NSvIR:Baryakhtar:DKHNS} and elsewhere, the mass capture rate is independent of $m_\chi$, hence the flat capture cross section for $m_\chi \gtrsim 0.5$~GeV. 
Below this mass, Pauli-blocking of final state neutrons by the Fermi-degenerate nucleon medium reduces the capture cross section in inverse proportion to $m_\chi$.
Below $m_\chi \simeq$ 35 MeV the sensitivities 
require further study as there is a new effect: in a superfluid NS, DM cannot scatter elastically on nucleons, since the DM kinetic energy is insufficient to excite nucleons above the $\Oc$(MeV) superfluid energy gap.
In principle low-mass DM can still capture in NSs via excitation of collective modes such as phonons (see, e.g. Ref.~\cite{NSvIR:Pasta}) and scattering with non-superfluid portions of the NS crust; we leave this to future work.

{\bf \em Infrequent NS encounters.}
For the scenario of $t_{\rm meet} > t_{\rm cool}$
we cannot expect a steady-state temperature. Thus the strategy must be to look for NSs not long after encounter -- corresponding to the right side of Fig.~\ref{fig:tvsTE} -- before they cool to temperatures below telescope sensitivities.
This is done by surveying the sky for anomalously hot NSs.

In Fig.~\ref{fig:$} top panel (again corresponding to the black capture region in the bottom panel) we show regions where at least 100 NSs with $\tilde T_s > 10^4$~K 
are expected in the ``solar vicinity," defined as a kpc-radius circle around the Sun. 
Note that for the range of $\Rsh$ displayed, the subhalo crossing time is less than $\tmeet$, where our treatment is valid. 
Currently, the HST is able to achieve point-source sensitivity as cold as $3\times 10^4$ K~\cite{coldestNSHST}.
Similar
directed observations of nearby pulsars with telescopes like LUVOIR~\cite{LUVOIR} may provide an alternative strategy to find anomalously hot NSs.
In addition, the observation of isolated $> 10^3$~K NSs may be undertaken (see Supplementary Material) by the Roman/WFIRST survey~\cite{RomanWFIRST} and in conjunction with deep-field observations by JWST, ELT and TMT \cite{NSvIR:Baryakhtar:DKHNS}.
We note that future optical survey sensitivities may detect $10^4$ K NSs within $\sim$20 parsecs, as discussed in the Supplementary Material. 

To obtain these regions we first estimate the probability of $\TNSz$ being at least some $\tilde T_s$, when heated by subhalo passage to a surface temperature $\tilde T^{\rm hot}_s$, as
\beq
p(\tilde T^{\rm hot}_s > \TNSz > \tilde T_s) = 1 - e^{- t_s/\tmeet}~,
\label{eq:probobsNS}
\eeq
where $ t_s = t_\mathrm{cool}(\tilde T_s)-t_\mathrm{cool}(\tilde T_s^{\rm hot})$ is the time taken for the NS to cool from $\tilde T^{\rm hot}_s$ to $\tilde T_s$;
see Supplementary Material for the derivation.
We then multiply this probability with the number of NSs $N^{\odot}_{\rm NS}$ in the solar vicinity obtained from the NS surface density of Model 1A* in Ref.~\cite{NSdistribs:Sartore2010}.
This model assumes a Maxwell distribution of NS velocities consistent with our encounter rate, and gives $N^{\odot}_{\rm NS} \simeq 2\times10^5$. 
As in Ref.~\cite{NSdistribs:Sartore2010} we assume that NSs are formed at a constant rate, so that the NS age distribution is peaked at the age of the Galaxy $\simeq 5 \times 10^9$~yr.\footnote{We have verified that including a broader distribution of NS ages (as given in Ref.~\cite{NSdistribs:Sartore2010}) does not noticeably change our results.}
This implies that the vast majority of the NSs we consider would have $\tilde T_s \ll 1000$~K in the absence of subhalo heating ($cf.$ the NS cooling curve in Fig.~\ref{fig:tvsTE}).

{\bf \em Other probes.}
C\acro{osmic} R\acro{ay} S\acro{cattering}.
DM particles with stronger-than-electroweak cross sections would undergo appreciable scatters with Galactic CRs, which has been used to bound DM's scattering cross section by studying the subsequent softening of the CR spectrum~\cite{reverseDD}, and
by constraining the flux of gamma-rays~\cite{CR:Cyburt:2002uw} and up-scattered DM~\cite{CR:BringmannPospelov,CR:Ema:2018bih,CR:NuExpCappiello:2019qsw,CR:PROSPECT:2021awi} 
produced in DM-CR collisions.
These limits also apply to subhalo DM, as the DM optical depth of CRs $\tau_{\rm CR}$ over long distances is unaffected by DM clustering.

In the top panel of Fig.~\ref{fig:$}, we show the $\Msh$-$\Rsh$ range constrained by CR scattering, for the $\sigma_{N\chi}$-$\mdm$ range shown in the bottom panel. The top panel region is bounded from below by requiring that $\tau_{\rm CR} > 1$ (setting the effective diffusion length = 8 kpc~\cite{BringmannCRgeometry,CR:BringmannPospelov}). 
The upper bound is simply the requirement that DM is appreciably clustered in subhalos.

P\acro{aleo}-D\acro{etectors}.
Since they contain Gyr timescale archeological records of particle interactions, ancient mineral slabs are potentially sensitive to the historic passage of DM subhalos through the Earth. 
Decades ago, Refs.~\cite{Price:1986ky,SnowdenIfft:1995ke} derived bounds on (unclumped) DM and monopole interactions by analyzing the microscopic structure of ancient mica. 
Reference~\cite{Baltz:1997dw} pointed out that ancient mica could be used to search for subhalo DM. Here, we recast bounds from Refs.~\cite{Price:1986ky,SnowdenIfft:1995ke}, using the fact that for some subhalo model space the total number of DM-mica scatters is unchanged by DM-clumping, 
discussed more in the Supplementary Material.
We thus show existing constraints in both panels of Fig.~\ref{fig:$} similar to CR-scattering limits, with the $\Rsh$-$\Msh$ region now bounded on the right by the requirement that the timescale of Earth encounters with DM subhalos is smaller than the age of the mica samples (determined from fission track dating~\cite{SnowdenIfft:1995ke} to be $(5 \pm 1) \times 10^8$~yr).
Recently a number of proposals have suggested further mineralogic DM searches~\cite{Drukier:2018pdy,Edwards:2018hcf,Baum:2018tfw,Baum:2019fqm,Ebadi:2021cte,Acevedo:2021tbl}, including for subhalo DM~\cite{Baum:2021chx}. 

For $m_\chi \sim 0.1-1~{\rm GeV}$ ($m_\chi \sim 5-10^5~{\rm GeV}$), if $\gtrsim 10^{-3}$ ($\gtrsim 10^{-10}$) of DM is diffuse, stronger DD bounds~\cite{CRESST:2019jnq,XENON:2019zpr,PandaX-4T:2021bab} than for CRs (mica) apply.

In Fig.~\ref{fig:$} we have also shown constraints from gravitational microlensing surveys~\cite{microlens:erosogle,microlens:subaru}, with the reach in radii limited from below by each survey's maximum Einstein radius, and reach in mass limited by observational cadences and finite source effects.
While these constraints are agnostic to the scattering properties of DM, our probes are seen to highly complement them, reaching much larger and lighter subhalos.   
We have also indicated the Schwarzschild radii corresponding to large $\Msh$; the region below this line is unphysical.
If a small fraction of DM remains unclumped, a possibility we have not explored in this study, direct detection searches may set bounds on the $\sigma_{n\chi}$-$\mdm$ plane; on the other hand, our bounds on the plane apply for any distribution of DM into smooth and clumped components.

{\bf \em Discussion.}
This {\emph{Letter}} charts the way forward for uncovering scattering interactions of subhalo DM.
We have shown that cosmic ray scattering and mineral detection already provide some sensitivity, while novel methods using NS temperatures already constrain subhalo DM with strong self-interactions.
In the future, subhalos over a vast range of masses and sizes may be uncovered by observations of either individual NSs or NSs in survey ensembles.
Interestingly, imminent limits on the DM scattering cross section for subhalo-bound DM could become {\em stronger} than those from direct detection searches on unclumped DM (see Fig.~\ref{fig:$} bottom panel and e.g., Ref.~\cite{PandaX-4T:2021bab}). 

Several new avenues await exploration.
While DM could heat old NSs to detectable brightness by imparting kinetic energy, greater NS temperatures and other interesting signals could arise if DM annihilations in the NS core are present.
For instance, if DM thermalizes slowly with the NS, annihilations may turn on late, so that there are two reheating periods in our ``infrequent encounter" regime.
DM annihilations following capture of subhalo DM could also produce detectable signals in a variety of celestial bodies~\cite{bradleyclumpSun,Leane:2020wob,Leane:2021ihh,Leane:2021tjj}.
While we have focused on DM-nucleon scattering,
some of our approaches also apply to DM that scatters on leptons.
The lepton content of NSs, though relatively small, provides substantial sensitivity to DM-electron/muon scattering~\cite{NSvIR:Bell2019:Leptophilic,NSvIR:GaraniHeeck:Muophilic,NSvIR:Riverside:LeptophilicShort,NSvIR:Riverside:LeptophilicLong,NSvIR:Bell:ImprovedLepton}.
Upscattering of DM with CR electrons should produce a detectable high-speed flux of DM, as with CR protons~\cite{CR:Ema:2018bih}. 
Exploring DM-electron scattering would therefore extend our results to even lighter DM masses than depicted in Fig.~\ref{fig:$}.
While for simplicity we had assumed uniform subhalo masses, more realistic mass distributions could be investigated. 

Finally, our NS heating mechanism via DM subhalo encounters provides, in addition to probing the neutron portal to beyond-the-SM scenarios~\cite{NSheat:DarkBary:McKeen:2020oyr,NSheat:Mirror:McKeen:2021jbh}, further motivation to imminent astronomical missions to make exciting discoveries in fundamental physics.

\section*{Acknowledgments}

We thank Raghuveer Garani for discussions on neutron superfluidity and Kevin Zhou for correspondence about NS-subhalo collisions.
B.J.K.\ thanks the Spanish Agencia Estatal de Investigaci\'on (AEI, MICIU) for the support to the Unidad de Excelencia Mar\'ia de Maeztu Instituto de F\'isica de Cantabria, ref. MDM-2017-0765.
The work of J.\,B. and N.\,R. is supported by the Natural Sciences and Engineering Research Council of Canada. 
Research at Perimeter Institute is supported in part by the Government of Canada through the Department of Innovation, Science and Economic Development Canada and by the Province of Ontario through the Ministry of Colleges and Universities.
TRIUMF receives federal funding via a contribution agreement with the National Research Council Canada.

\section*{Appendix on neutron star passive cooling}

The time for the NS to cool to $\Tcoolz$, in a process that is insensitive to the starting temperature, is given by~\cite{coolinganalytic:Ofengeim:2017cum} 
\begin{widetext}
\begin{equation}
\tcool (\tilde T_9)/{\rm yr} = \begin{cases}
t_{\rm env} = s_1^{-k} q^{-\gamma} \big[\big(1+ (s_1/q)^k \tilde T_9^{2-n} \big)^{-\gamma/k} - 1 \big], \ \Tcoolz > \tilde T_{\rm env}~, \\
t_{\rm env} + (3s_2)^{-1} (\tilde T_9^{-2} - \tilde T_{\rm env}^{-2}), \ \ \ \ \ \ \ \ \ \ \ \ \ \ \ \ \ \ \ \ \ \ \  \Tcoolz \leq \tilde T_{\rm env}~,
\end{cases}
\label{eq:tcoolvTfull}
\end{equation}
\end{widetext}
where $\tilde T_9 = \Tcoolz/(10^9~{\rm K})$,
$q = 2.75 \times 10^{-2}$,
$s_1 = 8.88 \times 10^{-6}$,
$s_2 = 8.35 \times 10^4$,
$k = (n-2)/(n-\alpha)$ and
$\gamma = (2-\alpha)/(n-\alpha)$ with
$\alpha = 2.2$ and
$n=8$.
The temperature $\tilde T_{\rm env} \simeq 4000~$K corresponds to the time after which the surface and internal NS temperatures equalize due to the thinning of the crustal outer envelope. 

We now provide some physics background to Eq.~\eqref{eq:tcoolvTfull}.
First recall that temperatures denoted with a tilde are in the frame of a distant observer.
During quiescent periods the NS temperature evolution is given by 
\beq
c_{\rm v}(\tilde T) \frac{d\tilde T}{dt} = - L_{\nu}^\infty (\tilde T) - L_\gamma^\infty (\tilde T)~,
\label{eq:dTdtfull}
\eeq
where the neutrino luminosity of the NS as measured by a distant observer of our benchmark NS is given by~\cite{heatcapacity:ReddyPageHorowitz:2016weq}
\beq
L_\nu^\infty (\tilde T) = 1.33 \times 10^{39}~{\rm J/yr}~\bigg(\frac{\tilde T}{10^9~{\rm K}} \bigg)^8~,
\eeq
applicable for slow/modified Urca processes, which we take to be the only neutrino cooling mechanism as prescribed by the ``minimal cooling" paradigm~\cite{coolingminimal:Page:2004fy}. 
These processes dominate the NS cooling down to $\tilde T = 10^8$~K before photon cooling takes over.
The luminosity of photon blackbody emission from the NS surface is:
\beq
L_\gamma^\infty (\tilde T_s) = 4 \pi (1+z)^2 R_{\rm NS}^2 \tilde T^4_s~.
\label{eq:blackbodylum}
\eeq
To solve Eq.~\eqref{eq:dTdtfull} we need a relation between $T_s$ and $T$, which depends on the composition of the outermost envelope, which acts as an insulating layer for temperatures above $\Oc(10^3)$~K.
For smaller temperatures the envelope becomes too thin for insulation, and the surface temperature reflects the internal temperature~\cite{cooling:Yakovlev:2004iq,coolingcatalogue:Potekhin:2020ttj}.
We assume a standard iron envelope for which we have the following relation at high temperatures~\cite{TbTs-Fe-Pethick1983,NSenvelope:Beznogov:2021ijc}:
\beq
T_s = 10^6~{\rm K} \bigg[\bigg( \frac{M_{\rm NS}}{1.5~M_\odot}\bigg) \cdot \bigg(\frac{10~{\rm km}}{R_{\rm NS}}\bigg)^2\bigg]^{1/4} \bigg[\frac{T}{9.43\times 10^7~{\rm K}}\bigg]^{0.55}~.
\label{eq:TsTbFe}
\eeq
We identify the thin-envelope regime by solving for $T_s = T$ in the above equation, giving $T_{\rm env} = 3908$~K. 
Below this temperature we model the $T_s-T$ relation by simply setting $T_s = T$. 

We now have all the ingredients to solve Eq.~\eqref{eq:dTdtfull}. 
We do this by following the prescription in Ref.~\cite{coolinganalytic:Ofengeim:2017cum}, suitably improving it to account for the thin-envelope $T_s-T$ relation, and obtain the analytic solution in Eq.~\eqref{eq:tcoolvTfull}.

The heat capacity of an NS depends sensitively on the superfluid and superconducting properties of NS nucleons.
It has long been recognized that nucleons could indeed form Cooper pairs in NS cores with an MeV energy gap, corresponding to a critical temperature $T_c \simeq 10^{10}$~K~\cite{sfluid:Yakovlev:1999sk,sfluid:PageReddyReview:2006ud,sfluid:Haskell:2017lkl}. 
Observational fits to cooling curves support this hypothesis~\cite{coolingcatalogue:Potekhin:2020ttj}.
The heat capacity of degenerate matter is set by particle-hole excitations close to the Fermi surface, thus the energy gap exponentially suppresses the nucleon contribution to the heat capacity for NS temperatures $\ll T_c$, the temperature zone we are interested in. 
Hence our NS heat capacity is set by the contribution from degenerate electrons (whose pairing-up critical temperature is $<$~100~K), and is given by Eq.~\eqref{eq:heatcapelec}, obtained from analytic fits for the electronic $c_V$ 
provided in Ref.~\cite{cvanalytic:Ofengeim:2017xxr}.
In general we expect $\Oc(10\%)$ variations to the above arising from the choice of the equation of state (EoS) for NS core matter and of the NS mass-radius configuration; in this work we have simply used Eq.~\eqref{eq:heatcapelec}.

Finally, we note that while in standard NS cooling scenarios, NSs are expected to reach $\sim 100-10^3$ K over billion year timescales \cite{cvanalytic:Ofengeim:2017xxr,cooling:Yakovlev:2004iq}, in some NS thermal evolution scenarios a temperature as high as $10^4$ K can persistent, if there is some additional source of NS heating like an anomalously large initial magnetic field strength \cite{Pagecooling2000} or an exceptionally fast initial rate of spin paired with rotochemical heating for certain NS EOS models \cite{Hamaguchi:2019oev}. However, in both these cases there is still the expectation that many late-stage NS will have temperatures below $\sim 10^3$ K.

\bibliography{refs}

\newpage

\begin{center}
 {\LARGE SUPPLEMENTARY MATERIAL}
\end{center}

\begin{figure*}
    \centering
    \includegraphics[width=\textwidth]{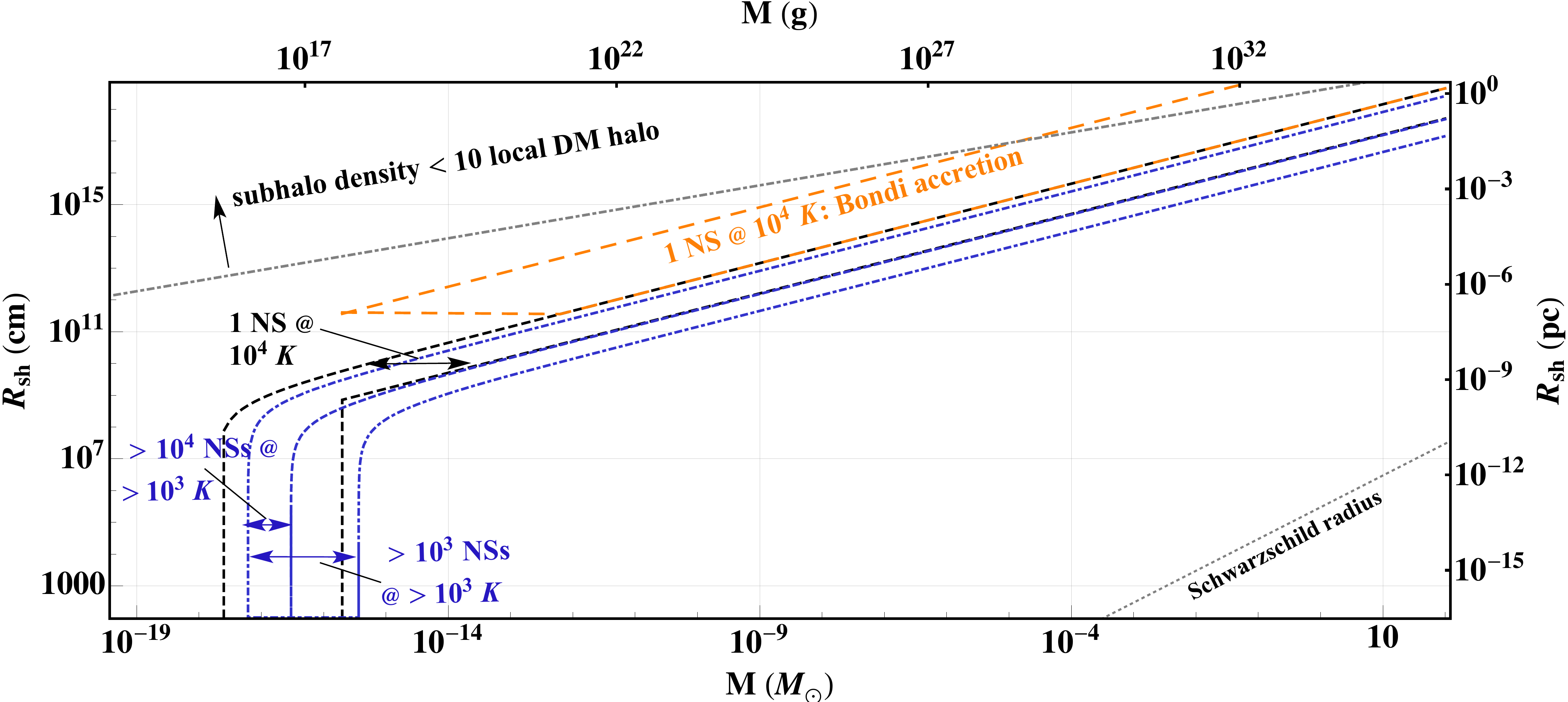}
    \caption{Same as the top panel of Figure~\ref{fig:$}, but showing the regions that may be constrained by observing
    (1) a single NS below 10$^4$~K; also shown is the region that may be probed if Bondi accretion of self-interacting subhalos is present,
    (2) at least 10$^3$ and 10$^4$ NSs below 10$^3$~K in an infrared survey.}
    \label{fig:SuppMatRvM}
\end{figure*}

\begin{figure}
    \centering
    \includegraphics[width=.49\textwidth]{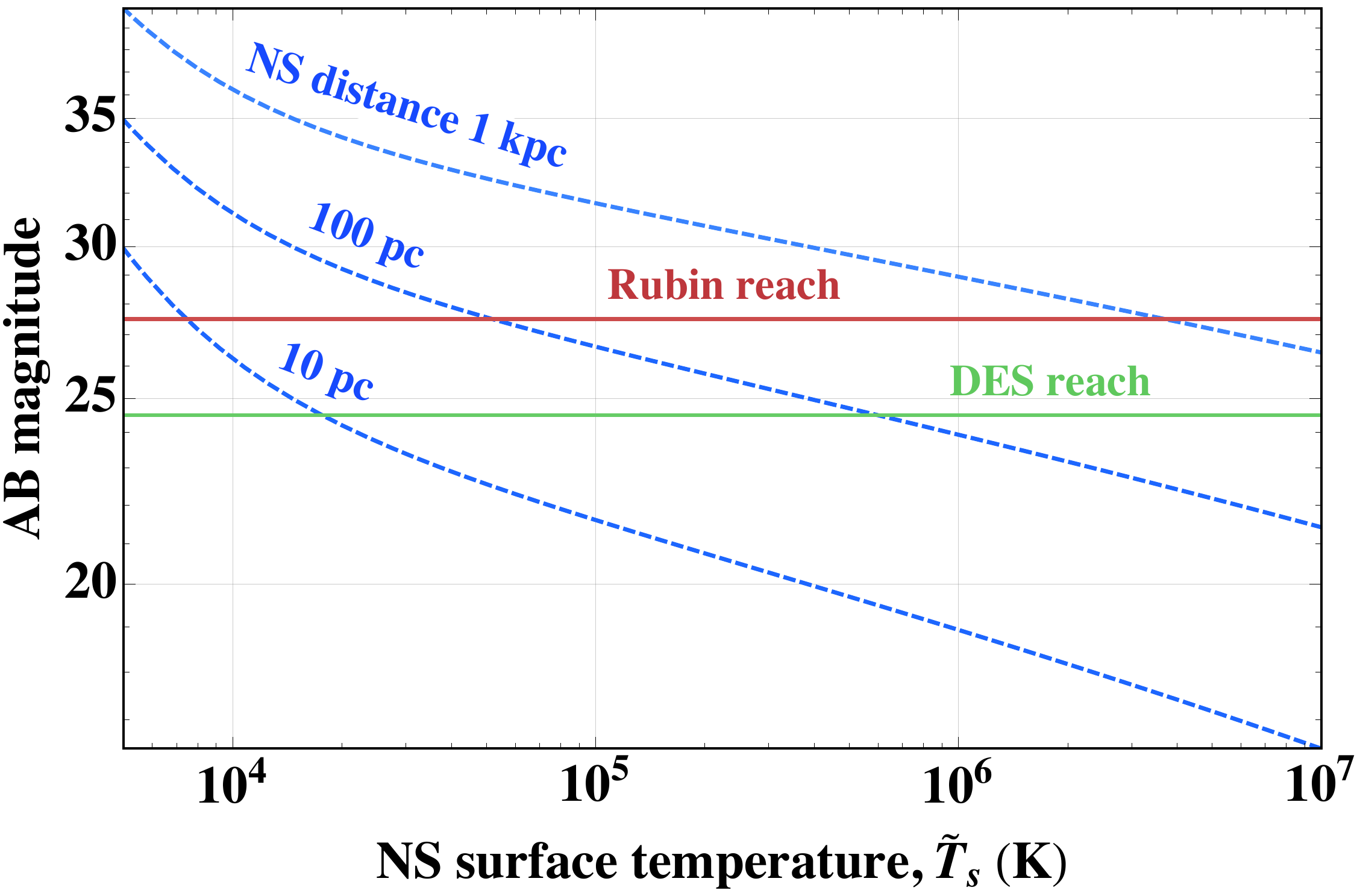}
    \caption{The g-band AB magnitude point source sensitivity of the deep field Dark Energy Survey and the projected Rubin sensitivity, compared to the luminosity of a neutron star with distances from Earth and surface temperatures indicated. Recall that larger magnitudes correspond to fainter objects.}
    \label{fig:SuppMatRubinDES}
\end{figure}

\section{Probabilities of observing heated neutron stars}

The probability of observing an NS above some temperature $\tilde T_s$, as given in Eq.~\eqref{eq:probobsNS}, is derived as follows.

Consider first $p_i(\tilde{T} > \tilde{T}_s)$, the probability that at the end of $i$ encounters the NS temperature $\tilde{T}$ at the time of observation  exceeds some threshold $\tilde{T}_s$. (We implicitly assume here that $\tilde{T}_\mathrm{hot} > \tilde{T}_s$.) 
We then note that $p_i(\tilde{T} > \tilde{T}_s)$ = $p_i(t < t_s)$, where $t$ is the time between the $i$th encounter and the observation (i.e.\ the time since the last encounter) and $t_s$ is the time taken to cool from $\tilde{T}_\mathrm{hot}$ to $\tilde{T}_s$. 
That is, if the most recent encounter occurred a time $t < t_s$ ago, then the NS temperature will still exceed $\tilde{T}_s$.

If there are exactly $i$ encounters over a time $t_\mathrm{NS}$, then
\bea
\nn    p_i(t < t_s) &=& 1 - p_i(t > t_s) \\
\nn            &=& 1 - p_i(N_{[t_\mathrm{NS} - t_s, t_\mathrm{NS}]} = 0) \\
            &=& 1 - \left( 1 - \frac{t_s}{t_\mathrm{NS}}\right)^i\,,
            \label{eq:p_i}
\eea
where $N_{[t_\mathrm{NS} - t_s, t_\mathrm{NS}]}$ is the number of encounters which occur in a time interval between $[t_\mathrm{NS} - t_s, t_\mathrm{NS}]$. 
The second term in the second line of Eq.~\eqref{eq:p_i} is therefore the probability of having no encounters within a time $t_s$ before the observation of the NS.
This is then evaluated explicitly in the next line by noting that the Poisson-distributed encounters occur uniformly over the time interval $t \in [0, t_\mathrm{NS}]$.

The expected number of encounters in a time $t_\mathrm{NS}$ is $\lambda = t_\mathrm{NS}/t_\mathrm{meet}$, hence the Poisson probability $p_{\mathrm{meet},i}$ of having $i$ encounters is
\beq
p_{\mathrm{meet},i} = \frac{\lambda^i e^{-\lambda}}{i!}~.
\label{eq:Poissonmeet}
\eeq

Accounting for multiple encounters, and using Eqs.~\eqref{eq:p_i} and \eqref{eq:Poissonmeet}, we obtain the required probability in Eq.~\eqref{eq:probobsNS} as
\begin{align}
    \begin{split}
        p(\tilde{T} > \tilde{T}_s) &= \sum_{i \geq 1} p_{\mathrm{meet},i} p_i(t < t_s)\\
        &= \sum_{i \geq 1} \frac{\lambda^i e^{-\lambda}}{i!} \left[ 1 - \left( 1 - \frac{t_s}{t_\mathrm{NS}}\right)^i \right] \\
         &= 1 - e^{-t_s/t_\mathrm{meet}}\,.
        \label{eq:p_T_s}
    \end{split}
\end{align}
Here we have included only terms with $i \geq 1$ because we assume that the equilibrium NS temperature exceeds $\tilde{T}_s$, so at least one encounter is required to be detectable.
Note that the above expression holds for all $t_s$ and $t_{\rm meet}$. 
In particular, for $t_s \ll t_{\rm meet}$, the exponential in Eq.~\eqref{eq:p_T_s} tends to zero and we end up with $p(\tilde{T} > \tilde{T}_s) \rightarrow 1$, as expected in the frequent encounter regime.

\section{NS-subhalo encounter cross section}

Here, we summarize the calculation of the gravitational encounter cross section for rigid objects and then extend to the case of non-rigid objects, such as the subhalos we consider in this work. This generalizes the brief discussion given in Ref.~\cite[p. 625]{BinneyTremaine:2008}.

Consider two objects of mass $m_1$ and $m_2$ approaching each other from infinity with relative velocity $v$ and impact parameter $b$. Conserving energy and angular momentum at infinity and at the distance of closest approach $r_\mathrm{min}$, the impact parameter can be written as:
\begin{equation}
    b^2 = r_\mathrm{min}^2 \left( 1 +  \frac{2 G (m_1 + m_2)}{v^2 r_\mathrm{min}} \right)\,.
\end{equation}
Two \textit{rigid} objects with radii $R_1$ and $R_2$ will collide if the distance of closest approach is $r_\mathrm{min} < R_1 + R_2$. This gives the well-known expression for the encounter cross section:
\begin{align}
\begin{split}
    \sigma_\mathrm{rigid} &= \pi b^2\\
    &= \pi (R_1 + R_2)^2 \left( 1 +  \frac{2 G (m_1 + m_2)}{v^2 (R_1 + R_2)} \right)\,,
\end{split}
\end{align}
as quoted in e.g.~Refs.~\cite{Kavanagh:2020gcy,Edwards:2020afl}.

\begin{figure}
    \centering
    \includegraphics[width=.49\textwidth]{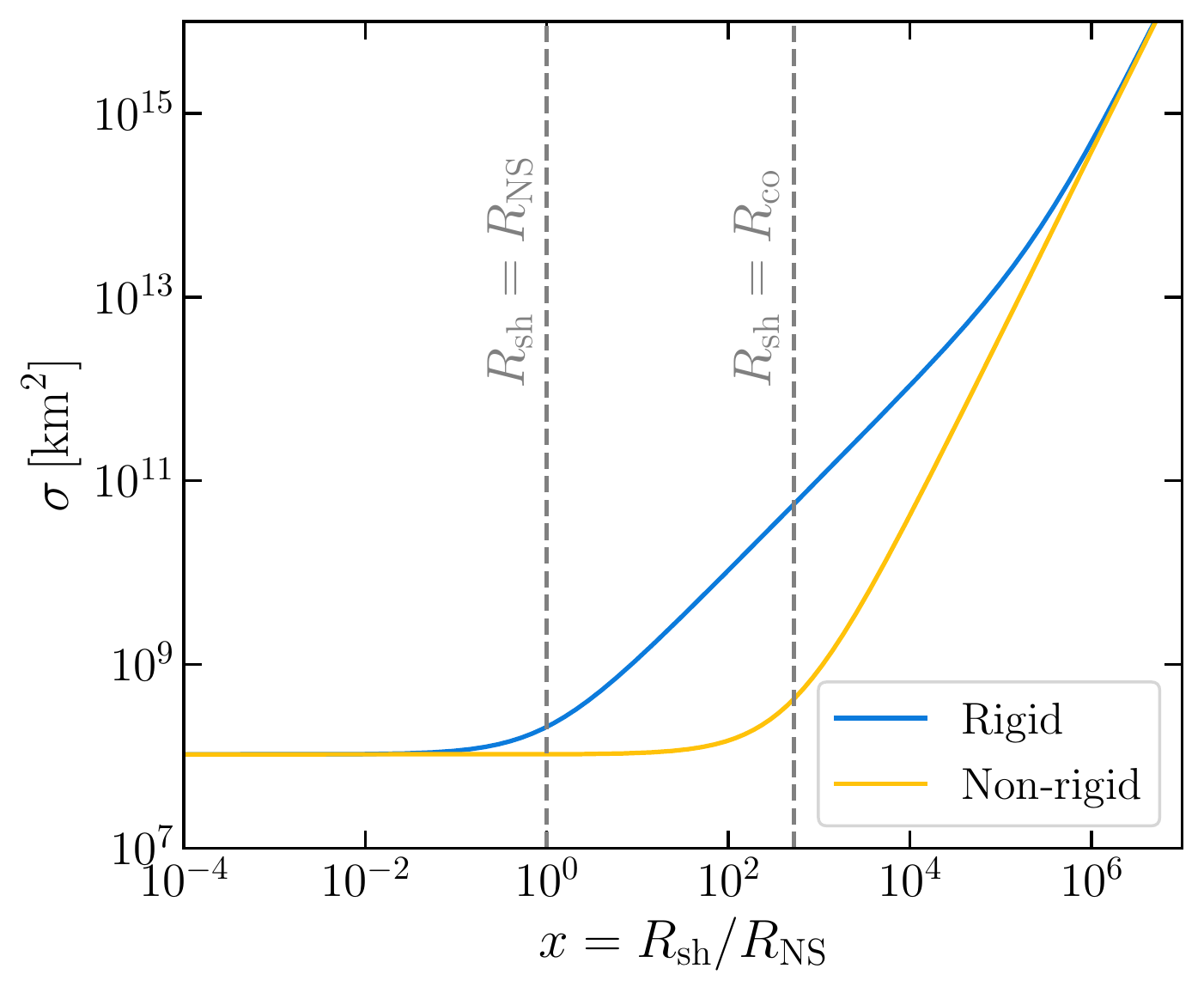}
    \caption{Cross sections for gravitational encounters with NSs. 
    The cross section for non-rigid subhalos (yellow) is visibly smaller than for rigid objects (blue) for $R_\mathrm{sh} \simeq R_\mathrm{co} = (\vesc/v)R_\mathrm{NS}$, and comparable for $\Rsh \ll R_{\rm co}$ and $\Rsh \gg R_{\rm co}$. 
    In both cases, we assume a relative velocity $v = 350\,\mathrm{km/s}$. The main text's analysis assumes non-rigid subhalos.}
    \label{fig:SuppMatCrossSections}
\end{figure}

However, in the case of NS-subhalo encounters, the NS may be considered rigid, but the subhalo is not. Instead, we should treat the particles in the subhalo as each being separately deflected by the NS. The subhalo will cross (at least some part of) the NS if the particles on the edge of the subhalo have $r_\mathrm{min} < R_\mathrm{NS}$; that is, if they lie a perpendicular distance closer than $b(r_\mathrm{min} = R_\mathrm{NS})$ from the NS. The required impact parameter $b'$ (measured with respect to the centre of the subhalo) is then obtained by adding the subhalo radius:
\begin{align}
\begin{split}
       b' &= R_\mathrm{sh} + b(r_\mathrm{min} = R_\mathrm{NS})\\ 
       &= R_\mathrm{sh} + \sqrt{R_\mathrm{NS}^2 + \frac{2 G M_\mathrm{NS}}{v^2 R_\mathrm{NS}}}\,,
\end{split}
\end{align}
where we have neglected the mass of the subhalo. The corresponding encounter cross section for a non-rigid subhalo is then:
\begin{align}
\begin{split}
        \sigma_\mathrm{Non-rigid} &= \pi (b')^2\\
        &= \pi\left(R_\mathrm{sh} + R_\mathrm{NS}\sqrt{1 + \vesc^2/v^2}\right)^2\,,
\end{split}
\end{align}
where we have defined 
$\vesc = \sqrt{2 G \MNS/\RNS}$. In Fig.~\ref{fig:SuppMatCrossSections}, we compare the rigid and non-rigid cross sections. The latter is always smaller, due to the fact that the non-rigid subhalo is `compressed' as it passes the NS. The difference is greatest for subhalos of size $R_\mathrm{sh} = R_\mathrm{co} = (\vesc/v)R_\mathrm{NS}$. Our analysis in this paper assumes non-rigid subhalos when computing DM capture on NSs.

\section{NS tidal effect on subhalo accretion}

As a subhalo passes near a NS, the tidal forces exerted by the NSs may cause the subhalo to pull apart and tidally disrupt~\cite{Rees1988,Dai:2021xcc}. 
For the case of collisionless DM subhalos, the amount of DM falling onto the NS will be the same as given in Eqs.~\eqref{eq:Gammameet}-\eqref{eq:rco}, since these expressions account for the gravitational trajectories of individual DM particles in the subhalos. However, in the case of DM subhalos with strong self-interactions, fluid properties of the subhalo can lead to circularized and disc-shape accretion flows onto NSs. 

To estimate what portions of DM subhalo parameter space shown in Figure~\ref{fig:$} might have accretion augmented in the case of fluid DM, we can compare the Roche limit of a NS for a subhalo of mass $M_{sh}$,
\bea
R_{\rm Roche} = R_{\rm sh} \left( \frac{2 M_{\rm NS}}{M} \right)^{1/3}
\eea
to the collection radius $R_{co}$ in Eq.~\eqref{eq:rco}. Equating these length scales, we find that for subhalos with radii larger than
\bea
R_{\rm sh} &\geq \bigg( \frac{\vesc}{\langle v_{\rm rel}\rangle} \bigg)^\ell \RNS (1 + z) \bigg(  \frac{M}{2 M_{\rm NS}} \bigg)^{1/3} \nonumber \\
& \gtrsim 2 \times 10^{11}~{\rm cm} \bigg( \frac{M}{M_{\odot}} \bigg)^{1/3},
\eea
it is possible that tidal disruption may cause circularized fluid DM accretion,
where this final expression has been normalized to our fiducial NS and halo parameters for the case of fluid DM accretion ($\ell=2$). However, it should be stressed that the above estimate does not guarantee tidal disruption, since the above estimate assumed a subhalo bound together only by gravity; stronger binding energies associated with a self-interacting DM subhalo would tend towards reducing tidal disruption. Assuming these subhalos are disrupted, they would form
circularized and disk accretion flows, whose accretion rates are an active topic of research. However, it is expected that tidal forces lead to an increase in the amount of material collected~\cite{Piran:2015gha}, compared to the spherical accretion of collisionless media.
Moreover, while our computation of the capture rate assumes that our subhalos are bound chiefly by gravity, again strong self-interactions and other internal dynamics could also be at play. 
If so, we see from the above treatment that the sensitivity to fluid DM clumps in Figure~\ref{fig:$} (region labelled PSR J2144) might be expanded.
We leave this possibility to future studies.

\section{Cosmic ray and paleolithic bounds on subhalo DM}

In order to adapt prior bounds on cosmic-ray upscattering to subhalo DM, we must ensure that cosmic rays that would usually scatter with diffuse DM, will equally scatter with subhalo DM. We can verify this condition by computing the cosmic ray optical depth to DM, $\tau_{\rm CR}$, for CRs contained in a diffusion zone of linear dimension $L_{\rm dfs}$ (which is around $1-15$ kpc~\cite{BringmannCRgeometry}) and subhalo number density $n_\chi = \rho_\chi/M$ as
\beq
\tau_{\rm CR} = \int_0^{L_{\rm dfs}} ds \ n_\chi(s) \sigma_{\rm geo} f_{\rm hit} = \frac{\rho_\odot}{M} \sigma_\mathrm{geo} 
\eeq
where the geometric cross section of the subhalo and the number of CR scatters within a subhalo (for per-nucleus cross section $\sigma_{N \chi}$) are given by  
\beq
\sigma_{\rm geo} = \pi R^2_{\rm sh}~, \ \ 
f_{\rm hit} = \frac{\sigma_{N \chi}}{\pi \Rsh^2} \frac{\Msh}{m_\chi }~.
\eeq
As indicated in the main text, $\tau_{\rm CR}> 1$ demarcates subhalo parameter space for which cosmic-ray upscattered DM bounds will apply.

Ancient mica searches for the smooth DM component assume that the DM particles will follow a Maxwell-Boltzmann distribution, with velocity dispersion $\sigma_v \sim 156\,\mathrm{km/s}$. However, the internal velocity dispersion of a light DM subhalo will be much smaller and so a DM clump would appear as a cold stream, with its velocity set by the relative encounter velocity of the Earth and the subhalo. Nonetheless, the encounter velocity will follow a similar Maxwell-Boltzmann-like distribution $f(v_\mathrm{rel})$, with typical values $v_\mathrm{rel} \sim 100 - 700\,\mathrm{km/s}$. As shown in Ref.~\cite{Ibarra:2019jac}, this stream-like velocity distribution for the subhalos is unlikely to alter the nuclear recoil rate by more than a factor of $\sim 2$. Furthermore, averaging over many subhalo encounters will wash out these differences, leading to a similar mean recoil rate as for the smooth distribution. We now demonstrate explictly that when averaging over a large number of subhalo encounters, the mean DM density experienced by a paleodetector is the same as in the smooth case, namely $f_\chi \rho_\chi$.

The rate of encounters depends on both the impact parameter $b$ and relative velocity $v_\mathrm{rel}$, so we can write:
\begin{equation}
    \frac{\mathrm{d}^2\Gamma_\mathrm{meet}}{\mathrm{d}b\mathrm{d}v_\mathrm{rel}} = \frac{f_\chi \rho_\chi}{M_\mathrm{sh}} \,2\pi b\, v_\mathrm{rel} f(v_\mathrm{rel})\,.
\end{equation}
During the encounter, the Earth crosses a distance $L = 2\sqrt{R_\mathrm{sh}^2 - b^2}$ across the subhalo, with a crossing time $t_\mathrm{cross} = L/v_\mathrm{rel}$. Averaging over the phase space of possible encounters, the mean DM density experienced by the Earth is:
\begin{align}
    \langle \rho_\chi \rangle &= \bar{\rho} \int\mathrm{d}b\,\mathrm{d}v_\mathrm{rel} \,t_\mathrm{cross} \frac{\mathrm{d}^2\Gamma_\mathrm{meet}}{\mathrm{d}b\mathrm{d}v_\mathrm{rel}}\,,
    \label{eq:meandensity}
\end{align}
where we assume that each subhalo has a uniform density $\bar{\rho} = (3 M_\mathrm{sh})/(4 \pi R_\mathrm{sh}^3)$. The integral in Eq.~\eqref{eq:meandensity} represents the fraction of time the Earth spends inside subhalos, and can be written explicitly as:
\begin{align}
    \langle \rho_\chi \rangle &= 4\pi\bar{\rho} \frac{f_\chi \rho_\chi}{M_\mathrm{sh}} \int_0^{R_\mathrm{sh}} b \sqrt{R_\mathrm{sh}^2 - b^2}\,\mathrm{d}b \int f(v_\mathrm{rel})\,\mathrm{d}v_\mathrm{rel}  \,.
\end{align}
The integral over the relative velocity distribution $v_\mathrm{rel}$ is normalised to unity, and the integral over $b$ evaluates to $R_\mathrm{sh}^3/3$. With this, the mean DM density experienced by a paleodetector on Earth (taking into account a realistic distribution of encounters) is then $\langle \rho_\chi \rangle = f_\mathrm{\chi}\rho_\chi$, the same as if the DM was smoothly distributed.

In order for existing constraints from ancient mica to be valid, then, we require that the encounter timescale (i.e.~the typical time between Earth-subhalo encounters) should be shorter than the age of the mica samples: $\tau_\mathrm{meet} \lesssim 5 \times 10^8\,\mathrm{yr}$. We note that the calculation above will not be strictly valid for such large values of $\tau_\mathrm{meet}$, because this would correspond to only $\mathcal{O}(1)$ encounter during the age of the sample. This means that close to the brown line in Fig.~\ref{fig:$}, the limits we proposed on subhalo DM are only indicative of the sensitivity of paleodetectors. However, as we increase $R_\mathrm{sh}$ away from the brown line the average over many encounters becomes an increasingly good approximation.

The encounter timescale is given by:
\begin{equation}
    \tau_\mathrm{meet} = \Gamma_\mathrm{meet}^{-1} = \left[\frac{f_\chi \rho_\chi(r_\odot)}{M_\mathrm{sh}} \sigma_\mathrm{geo} \langle v_\mathrm{rel}\rangle\right]^{-1}\,,
\end{equation}
neglecting any gravitational focusing effects from the Earth. Of course, a rare, glancing encounter with a massive subhalo, with impact parameter $b \approx R_\mathrm{sh}$, will lead to a much smaller signal in a paleolithic detector than a more head-on encounter, in which the Earth traverses more of the sub-halo. When estimating the region of validity of the ancient mica constraints, we therefore require -- in the calculation of $\tau_\mathrm{meet}$ -- that the encounter takes place at an impact parameter $b < \langle b \rangle = (2/3)R_\mathrm{sh}$. This corresponds to replacing $\sigma_\mathrm{geo} \rightarrow (4/9)\sigma_\mathrm{geo}$, and corrects for those scenarios where only a very small number of glancing encounters are expected, which would lead to a deficit of signal in paleo-detectors.

\section{More target regions for future telescopes}

In Fig.~\ref{fig:SuppMatRvM} we show regions further to those shown in Fig.~\ref{fig:$} that could be probed by future telescopes.
We see that the triangular region corresponding to observing a single NS at or below 10$^4$~K is consistent with Fig.~\ref{fig:tvsTE}.
We also see that in the presence of Bondi accretion (for subhalos with dissipative self-interactions), due to a larger collection radius than for collisionless DM (Eq.~\eqref{eq:EpassMsh}), a much larger $\Msh$-$\Rsh$ region may be probed for $\Rsh$ exceeding the Bondi radius.

In the coming decades, astronomical surveys have interesting prospects for the discovery of subhalo-heated NSs.  
In Fig.~\ref{fig:SuppMatRubinDES} we show the sensitivity of optical surveys alongside the luminosity of nearby NSs.
The Dark Energy Survey~\cite{DES} and the planned Rubin/LSST~\cite{RubinLSST,LSST:2008ijt} achieve g-band AB magnitude sensitivities of 24.5 and 27.5, compared with a fainter $31$ AB magnitude for a $10^4$ K NS at a distance of 0.1 kpc. We see that it may be possible for Rubin to discover or exclude the existence of subhalo-heated NSs within tens of parsecs from Earth.
This would provide some information on whether the Earth is located inside or near a subhalo.

Finally, for the sake of completeness we make a few comments about x-ray survey sensitivity to subhalo-heated NSs.
Using the ROSAT all-sky survey as well as infrared and radio source maps, limits have been set on the number of x-ray emitting NSs near the solar position~\cite{Rutledge:2003kg,Turner:2010pk}. 
The ROSAT flux count sensitivity of $5 \times 10^{-13} ~{\rm cm^{-2}~s^{-1}}$ in the $0.1 - 2.4$ keV band corresponds to an AB magnitude sensitivity of $25-28$, which has permitted the discovery of $\gtrsim 5 \times 10^5~\rm{K}$ NSs within $\sim 0.5$ kpc~\cite{Voges:1999ju}.
At present, ROSAT results limit the number of nearby x-ray emitting isolated NSs to less than 48 at 90\% C.L.~\cite{Turner:2010pk}.
The upcoming eRosita all-sky survey is expected to improve on the flux sensitivity by a factor of 25~\cite{Erosita:Predehl:2020waz}.
However, we find that the NS temperatures required for x-ray detection of NSs ($\gtrsim 10^5~{\rm K}$) lie well above temperatures that would be imparted by subhalo DM near the solar position. 
Nevertheless,  in future studies it would be interesting to explore the subhalo DM sensitivity that would be provided by x-ray searches for NSs near the Galactic Center, where DM densities are higher.

\end{document}